\documentclass{article}

\usepackage{epsfig}
\usepackage{amsmath}

\begin{document}

\numberwithin{equation}{section}

\title{Comparative Analysis of GALLEX-GNO Solar Neutrino Data and 
       SOHO/MDI Helioseismology Data; Further Evidence for 
       Rotational Modulation of the Solar Neutrino Flux}
\author{Peter A. Sturrock \& Mark A. Weber\\
        Center for Space Science and Astrophysics\\
        Stanford University, Stanford CA 94305}
\maketitle

\begin{abstract}

If the solar neutrino flux is variable, the variability could be caused
by modulation of the neutrino flux by an inhomogeneous magnetic field
in the solar interior. This modulation, which requires nonstandard
physics involving a nonzero neutrino magnetic moment, could lead to an
oscillation in the detected neutrino flux with a frequency set by the
synodic rotation frequency in the region of the solar interior that
contains the magnetic structure. We investigate this possibility by
carrying out a comparative analysis of the GALLEX-GNO solar neutrino
data and estimates of the solar internal rotation rate derived from the
MDI helioseismology experiment on the SOHO spacecraft.

We first carry out a Lomb-Scargle analysis of the data. There is no
significant peak in the spectrum that could be associated with the
radiative zone. On the other hand, there are peaks that could be
associated with the convection zone. We introduce a statistic, which we
evaluate as a function of radius and latitude, that is a measure of the
degree of ``resonance'' of oscillations in the neutrino flux and the
synodic solar rotation rate at that radius and latitude. If the
oscillations are real and caused by an internal magnetic structure, the
resonance statistic will indicate the probable location of that
structure. A map of this statistic indicates that the probable location
is deep in the convection zone near the equator.

We next modify this statistic by integrating it over the equatorial
section of the convection zone. This integral resonance statistic
provides a measure of the likelihood that the variability of the solar
neutrino flux, as measured by the GALLEX-GNO data, has its origin in
the equatorial section of the convection zone. We apply the shuffle
test, randomly reassigning measurements among runs, to estimate the
significance of the value of the statistic computed from the actual
data. This test implies that the result is significant at the 0.2\%
level. When, for comparison, we repeat this analysis for the radiative
zone, we find that the integral resonance statistic is not
significant.  These results support earlier evidence for rotational
modulation of the solar neutrino flux.

\vspace{5mm}
\noindent {\em Subject headings:}~Sun: interior---Sun: particle
                        emission---Sun---rotation

\end{abstract}

\section{Introduction}

There has for some time been great interest in the possibility that the
solar neutrino flux is variable. Early tests focused on the possibility
that the flux varies with the solar cycle.  This period (11 years) is
so long that power spectrum analysis is not feasible. Hence the
preferred test for variability was an indirect one, namely, the search
for a correlation between the solar neutrino flux and an index of solar
variability such as the Wolf sunspot number (Bahcall, Field \& Press
1987; Bahcall \& Press 1991; Bieber et al. 1990; Dorman \& Wolfendale
1991), the surface magnetic field strength (Massetti \& Storini 1993;
Oakley et al.\ 1994), the intensity of the green-line corona (Massetti
\& Storini 1996), or the solar wind flux (McNutt 1995).  Walther (1997)
has criticized such claims---specifically the claim that the neutrino
flux is anti-correlated with the sunspot number---on the grounds that
the authors used smoothed data and then applied tests that are
appropriate only if data points are independent. Snodgrass and Oakley
(1999) have in turn criticized Walther's article, and Walther (1999)
has responded to that criticism. The implications of correlation
analyses remain in doubt.

Although the overall duration of solar neutrino data (24 years for the
Homestake experiment) is not long enough to permit a definitive
spectrum analysis for oscillations as slow as the solar cycle, the
duration is adequate for a search for oscillations of shorter period.
Haubold and Gerth (1990) have used spectrum analysis to search for
evidence of oscillations with period of order one year. More recently,
Haubold (1998) has used wavelet analysis for this purpose.

It is well known that indices such as sunspot number, surface magnetic
field strength, coronal brightness, etc., are strongly influenced by
solar rotation.  They display oscillations with a basic synodic period
of about 27 days and harmonics of this frequency.  These oscillations
arise from the fact that the solar magnetic field is highly
inhomogeneous, and that some components of this field last for several
or many solar rotations.  Neugebauer et al.\ (2000) have recently
presented evidence of a rotational pattern in the solar wind that lasts
for more than three solar cycles. Compared with the duration of
neutrino experiments, the solar rotation period is short enough that it
is quite reasonable to search for the influence of rotation by means of
time-series spectrum analysis.

According to ``standard physics,'' one would not expect the solar
neutrino flux to vary. However, it has been realized for some time that
neutrino processes may be governed by ``nonstandard physics'' (see, for
instance, Raffelt 1996). It appears that the currently most popular
candidate for the solution of the solar neutrino deficit (see, for
instance, Bahcall, Krastev \& Smirnov 1998) is the MSW effect (Mikheyev
\& Smirnov 1986a, 1986b, 1986c; Wolfenstein 1978, 1979), whereby
electron neutrinos may be converted into either muon or tau neutrinos
as they propagate through matter in the solar interior. The possibility
that neutrino magnetic moment may have something to do with the
neutrino deficit was first advanced by Cisneros (1971), who considered
propagation through magnetic field in the core of the Sun. This leads
to spin precession (see also Fujikawa \& Schrock 1980), which converts
some of the left-handed electron neutrinos, that are produced by
nuclear reactions in the core, into sterile right-hand neutrinos which
are not detectable. At a later date, Voloshin, Vysotskii, and Okun
(1986a, 1986b) and Barbieri and Fiorentini (1988) examined the possible
variation of the solar neutrino flux due to propagation through the
solar convection zone, taking account of the effect of matter as well
as magnetic field. They found that matter tends to suppress spin
precession, although spin precession is still possible if the product
of the magnetic moment times the magnetic field strength is large
enough. This is now known as the ``VVO'' effect. Schechter and Valle
(1981, 1982) considered the effect of a possible flavor-off-diagonal
(transition) magnetic moment, and found that this could lead to the
simultaneous precession of both spin and flavor (``spin-flavor
precession'' or ``SFP'' if neutrinos propagate through a magnetic
field. Akhmedov (1988a,b) and Lim \& Marciano (1988) analyzed the
propagation of neutrinos through matter permeated by magnetic field ,
with application to propagation through the solar convection zone. They
found that, for a given neutrino energy, there is a certain density at
which a resonant process occurs, enhancing spin-flavor precession.
This is now known as ``resonant spin-flavor precession'' (RSFP).  The
implications of RSFP for solar neutrinos have been reviewed more
recently by Akhmedov (1997) and Pulido \& Akhmedov (2000).

Much of the above analysis was stimulated by the possibility that the
solar neutrino flux may vary in the course of the solar cycle, although
Voloshin, Vysotskii, \& Okun (1986a, 1986b) also drew attention to the
possibility that the flux may vary in the course of a year, due to a
possible latitude-dependence of the internal magnetic field and to the
inclination of the Sun's rotation axis to the ecliptic. These issues
remain important but, for reasons given above, we have become more
interested in the possibility that the flux may vary with the solar
rotation frequency. Such an effect is most likely to show up if the Sun
contains strong, extensive and long-lived magnetic structures,
especially if they are situated where the Sun exhibits something close
to rigid rotation.

The solar interior is divided into two sharply different region by the
``tacho\-cline,'' located at about $r = 0.7$. (Here and elsewhere the
radius is normalized to that of the photosphere.) Nuclear burning
occurs near the center: 80\% of the burning occurs within $r = 0.15$
(see, for instance, Bahcall 1989, p.~147). Between the nuclear-burning
core and the tachocline is the radiative zone.  According to data
derived from helioseismology (Schou et al.\ 1998), the core and
radiative zone are in substantially rigid rotation with a sidereal
rotation rate in the range $13.75 \pm 0.25 \,{\rm y}^{-1}$. We find it
convenient to measure frequencies in cycles per year, since this leads
to a simple relationship between the sidereal and synodic values:
\begin{equation}
\nu({\rm synodic}) = \nu({\rm sidereal}) - 1 .
\end{equation}

At the tachocline, the sidereal rotation rate jumps from about $13.75
\,{\rm y}^{-1}$ at $r = 0.66$ to about $14.53 \,{\rm y}^{-1}$ at $r =
0.74$. Above the tachocline, in the convection zone which extends to
the photosphere, the rotation rate varies with radius and latitude,
from a minimum of about $8.68 \,{\rm y}^{-1}$ at $r = 1$ at the poles to
a maximum of about $14.84 \,{\rm y}^{-1}$ at $r = 0.93$ at the equator.
Early in our research, these considerations led us to focus our
attention on the radiative zone since this is larger than the
convection zone, contains gas of higher pressure that could contain
stronger magnetic field, and is believed to rotate effectively like a
rigid body.

In searching for evidence for rotational modulation of the solar
neutrino flux by examining radiochemical data, one is faced with the
serious difficulty that the timing of the data acquisition is
comparable with the period of oscillations one is looking for. These
data have been acquired in runs, each lasting two to four weeks, the
runs being spaced at intervals of two to four weeks. Aware of this
difficulty, we nevertheless proceeded to search for an oscillation in
the Homestake data with a frequency comparable with the synodic
rotation frequency of the radiative zone ($12.75 \pm 0.25
\,{\rm y}^{-1}$), using a maximum-likelihood procedure (Sturrock et al.\
1997). We found a strong peak at $12.85 \,{\rm y}^{-1}$. We have also
found some evidence that the solar neutrino flux, as measured by the
Homestake experiment, exhibits modulation related to heliographic
latitude (Sturrock et al.\ 1998).

The GALLEX (Anselmann et al.\ 1993, 1995; Hampel et al.\ 1996, 1997)
and GNO (Altman et al.\ 2000) consortia, working with the same gallium
experiment, have also acquired extensive radiochemical measurements of
the solar neutrino flux. We have published a preliminary spectrum
analysis of the GALLEX data (Sturrock et al.\ 1999), based on a
least-squares procedure (Knight, Schatten, \& Sturrock 1979), which
shows some evidence of a periodicity that may be related to solar
rotation.

As an independent investigation of the question of variability, we have
recently studied the histogram of GALLEX-GNO data (Sturrock \& Scargle
2001). This histogram certainly appears to be bimodal, but this claim
needs further investigation. A bimodal histogram would be incompatible
with a stationary time series but could be compatible with a neutrino
flux with a large-amplitude oscillation with a period of order a few
weeks.

The timing of the radiochemical experiments is (fortunately) not
completely regular. Bretthorst (1988) has shown by Bayesian analysis
that the Lomb-Scargle technique of spectrum analysis (Lomb 1976,
Scargle 1982) is the optimum procedure to apply to irregularly spaced
data. He has shown that irregular timing, far from being a drawback, is
in fact a great asset, permitting finer frequency resolution and
mitigating the adverse effect of aliasing. We present a Lomb-Scargle
spectrum analysis of the GALLEX-GNO data in Section 2. We do not find a
single dominant peak in the band of frequencies that may be associated
with the radiative zone. However, we do find more than one peak in the
frequency band associated with the convection zone.

In Section 3, we consider the solar interior on a point-by-point
basis, introducing a statistic that measures the degree of resonance
between the oscillation of the neutrino-flux measurements and the local
rotation. If one were to assume that the oscillation is real and is due
to modulation by a magnetic structure (or structures) in the solar
interior, this map would indicate the probable location of the
structure. We find, from examination of the map, that the modulation
(if real) probably occurs in the low-latitude region of the lower half
of the convection zone. This finding is compatible with our theoretical
expectation that modulation should occur near the equator since, as we
discuss in Section 4, neutrinos detected on Earth must have passed
through the convection zone in a small band of latitudes.

In Section 4, we carry out a statistical evaluation of the significance
of the above result. By integrating the point-by-point resonance
statistic over the equatorial section of the convection zone, we form
an ``integral resonance statistic'' that is a measure of the
correlation between the structure of the power spectrum and the
distribution of rotation rates in the entire convection zone. We
compare the actual value of this statistic with the distribution of
values found from simulated (``shuffled'' data obtained by randomly
reassigning the measurements among runs. Of 10,000 simulations, only 16
have values larger than the actual value, indicating that the
correspondence between the power spectrum and rotation in the
convection zone is significant at the 0.2\% level.

For comparison with the above result, we present in Section 5 a similar
comparative analysis of the GALLEX-GNO spectrum and the rotation
profile of the radiative zone. The shuffle test shows no evidence that
the solar neutrino flux is modulated by a structure or structures
embedded in the radiative zone.

In Section 6, we comment upon the findings of this article and indicate
some of the follow-up research that will be required.

\section{Lomb-Scargle Analysis}

We here present the results of a Lomb-Scargle spectrum analysis of the
GALLEX-GNO data.  This analysis will be discussed in more detail in a
later article, where we will evaluate the strengths and weaknesses of
the procedure as applied to neutrino data, and compare of spectra
obtained from GALLEX-GNO data and from Homestake data.

Most of the runs last about four weeks but, on the other hand, the
half-life of the capture product $^{71}{\rm Ge}$ is only 11.43 days
(Bahcall 1989, p.~343).  The survival probability of a germanium
nucleus produced on the first day of a run is only 18\%.  Hence the
measurement of each run is weighted towards captures late in the run.
We shall show in our forthcoming article that, when we calculate the
amplitude of the oscillation of the flux measurements caused by a
sinusoidal modulation of the neutrino flux with periods of order four
weeks, the measurement of each run is even more heavily weighted in
favor of events late in the run.  For this reason, we can obtain a
satisfactory estimate of the spectrum of the neutrino flux by assigning
the measurement made during each run to the end time of that run. This
is not to minimize the challenge posed by the goal of extracting the
power spectrum of the neutrino flux from measurements acquired from
runs lasting three or four weeks.  These difficulties will be discussed
in our later article. As Bretthorst (1988) has shown, it is a distinct
advantage that the timing of data we are analyzing has not been
completely regular (although we could have wished the timing to have
been more irregular than it actually is).

The result of our Lomb-Scargle analysis of the GALLEX-GNO data is shown
in Figure 1, for frequencies in the range 10 to $20 \,{\rm y}^{-1}$.  If
modulation were occurring in the radiative zone, we might expect a
fairly well defined peak, since that region is believed to be in rigid
rotation.  According to helioseismology data (Schou et al.\ 1998), the
rotation rate of the radiative zone is in the range 12.5 to $13.0 \,{\rm
y}^{-1}$, and probably in the range 12.6 to $12.9 \,{\rm y}^{-1}$.  We
see from Figure 1 that there is no significant peak in this range.
There is an interesting peak with $S = 4.5$ at $\nu = 13.04 \,{\rm
y}^{-1}$ which, if real, may be related to the tachocline.

\begin{figure}[t]
\epsfig{file=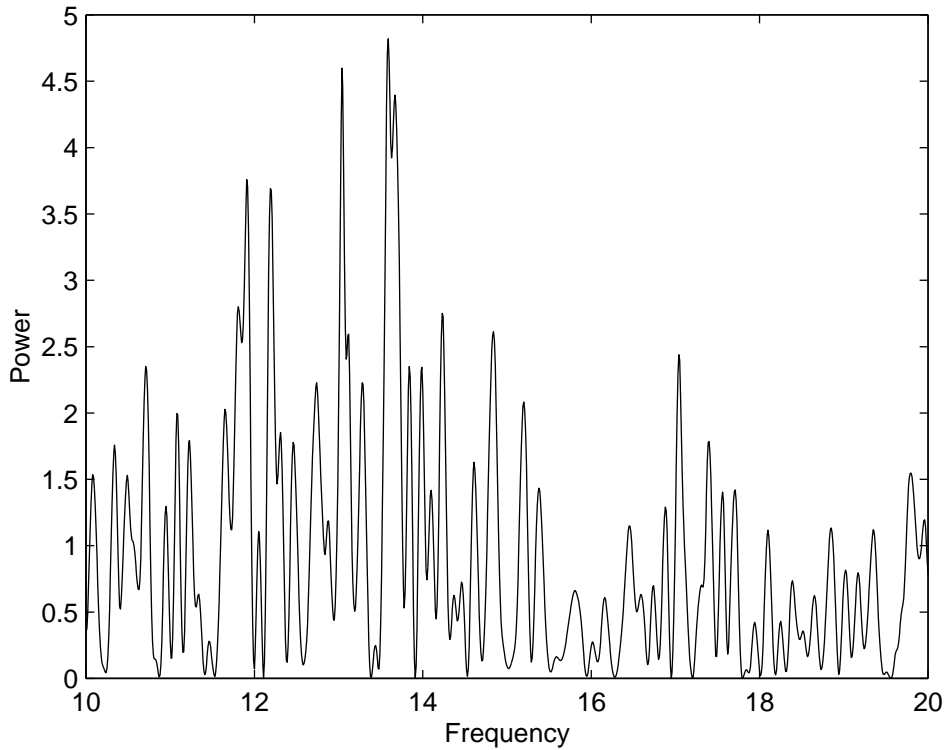}
\caption{Lomb-Scargle spectrum of the GALLEX-GNO solar neutrino 
         data for the frequency range 10--20 ${\rm y}^{-1}$.}
\end{figure}

The rotation rate of the equatorial section of the convection zone is
in the range 14.3 to $14.8 \,{\rm y}^{-1}$ sidereal or 13.3 to $13.8 \,{\rm
y}^{-1}$ synodic.  This band represents a combination of uncertainty in
the measurements and a real variation of the rotation rate with
radius.  We see, in Figure 1, that there is a prominent double peak in
this band, each peak having power of about $S=4.5$.  For a given
frequency, the probability of finding a peak of power $S$ or more is
$e^{-S}$ (Scargle 1982).  If there are indeed two independent peaks,
the combined significance would be notable.

\section{Map of Resonance Statistic}

Our goal is to examine the relationship between the variability of the
neutrino flux and internal rotation. It is convenient to begin by
introducing a visual display of this relationship. Schou et al.\ (1998)
have tabulated the rotation rate $\nu_h(r,\lambda )$ and the error
estimate $\sigma_h(r,\lambda )$ for 101 values of the radius $r$ and 25
values of the latitude $\lambda$ These data determine a probability
distribution function (PDF) of the rotation frequency for each pair of
values $r, \lambda$:
\begin{equation}
P(\nu|r,\lambda ) = (2\pi)^{-1/2} \sigma_h\;^{-1}
               \exp\left[-\frac{(\nu - \nu_h)^2}{2\sigma_h\;^2}\right] .
\end{equation}

Since we wish to compare the rotation frequencies with the
neutrino-flux variability as measured on Earth, it is appropriate to
inspect the synodic rotation rates rather than the sidereal rates.

We can define a measure of the degree of resonance of the neutrino
flux with internal rotation by forming the following ``resonance
statistic,''
\begin{equation}
\Xi(r,\lambda ) = \int\limits_{\nu_a}^{\nu_b} {\rm d}\nu S(\nu) 
                  P(\nu|r,\lambda ) ,
\end{equation}
where $S$ is the Lomb-Scargle spectrum computed in Section 2. In this
integral, we need to select limits of integration that are sufficiently
wide to cover all significant contributions from the PDF. We have
adopted $\nu_a = 0$ and $\nu_b = 20$, but a much smaller range would
have been satisfactory. A map of $\Xi$ as a function of radius and
latitude is presented in Figure 2.

\begin{figure}[p]
\epsfig{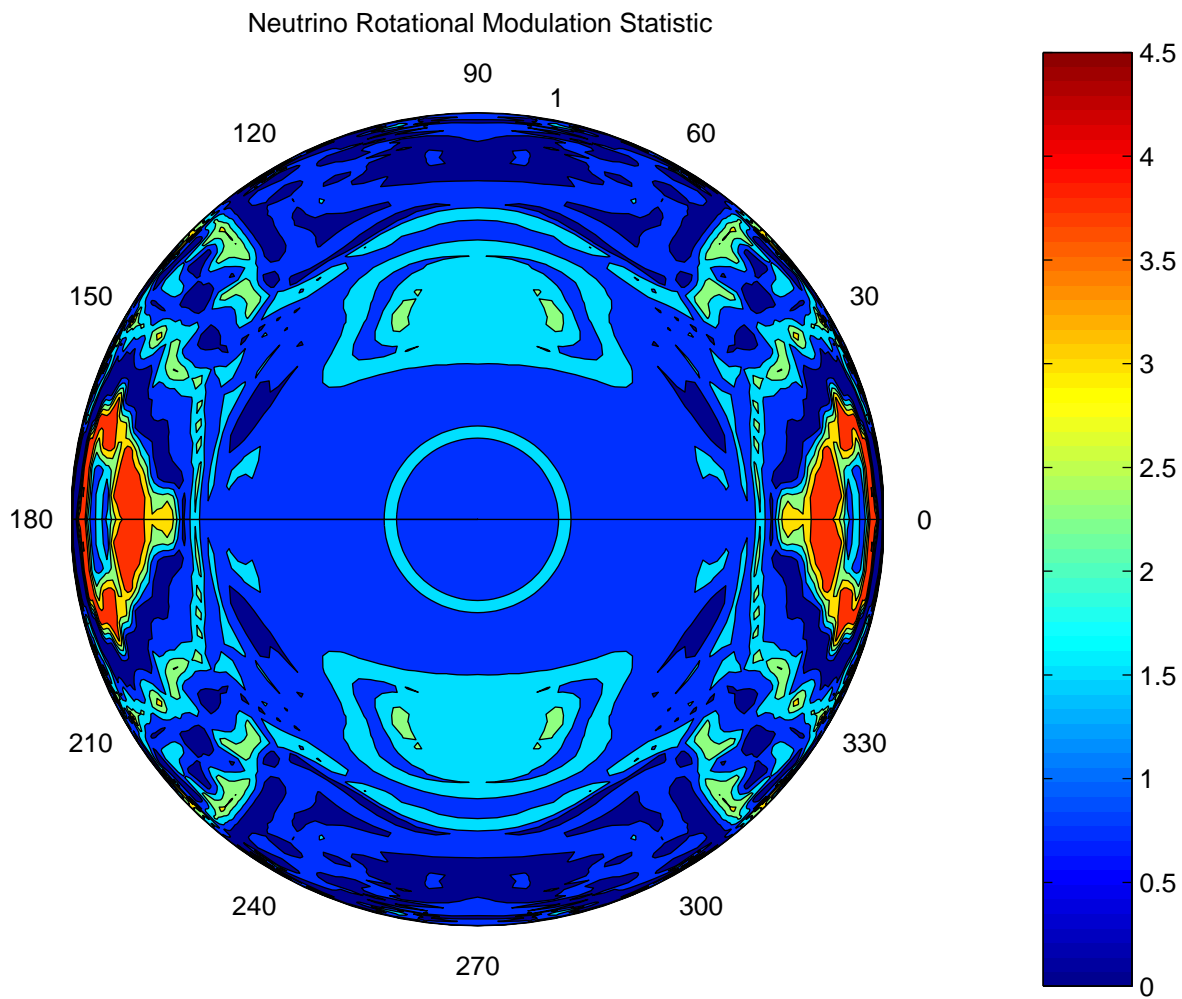}
\caption{Map of the resonance statistic $\Xi$, defined by equation (3.2), 
         upon a meridional section of the solar interior. Red or yellow 
         denotes a "resonance" between the neutrino flux and the local 
         solar rotation, indicating that the two oscillations (flux 
         variability and rotation) have the same frequency.}
\end{figure}

This figure is essentially a mapping of the power spectrum of the
solar-neutrino time series onto the solar interior. Where the map is
colored yellow or red, $\Xi$ is large compared with its average value
of unity. This denotes a ``resonance'' between the neutrino flux and
the local solar rotation in the sense that the two ``oscillations''
(flux variability and rotation) have the same frequency. When the color
is blue, there is no such resonance. If two regions of the solar
interior have the same rotation rate, they will tend to have the same
value of the statistic $\Xi$.

Another way of looking at this figure, which is perhaps physically more
significant, is the following:- Let us assume that there is a
well-defined oscillation in the neutrino flux, and let us assume that
this oscillation is due to modulation of the flux by a structure (such
as a magnetic structure) in the solar interior. Then we can attempt to
locate that structure by finding the location (or locations) where the
rotation rate has just the correct value to account for the dominant
oscillation of the neutrino flux. The map shown in Figure 2 may now be
viewed as a PDF for the location of the modulating structure.

We see from the map that, if the neutrino flux is variable, and if the
variability is due to modulation by some internal solar structure, then
that structure is probably in the lower part of the convection zone and
(as we would expect) near the equator. The map offers no evidence that
such modulation is occurring in the core or in the radiative zone.

\section{Significance of the Apparent Convection-Zone Modulation}

We now need a statistical evaluation of the
significance of the modulation of the neutrino flux by a structure or
structures within the convection zone that is suggested by Figure 2.

Most neutrinos detected by the GALLEX-GNO experiment are produced by pp
(proton-proton) reactions, and 80\% of the pp reactions occur within a
radius of 0.15 of the center of the Sun (see, for instance, Bahcall
1989, p.~147). Near the surface of the Sun, this corresponds to a range
$\pm 8.5$ degrees of heliographic latitude.  We also note that the
Sun's axis is tilted at about 7.25 degrees with respect to the
ecliptic.  Hence most of the neutrinos detected by GALLEX-GNO have
passed well within 16 degrees of the solar equator in penetrating the
surface of the Sun. Over the maximum range of 16 degrees, the rotation
rate varies by less than 1 percent within the convection zone.  This
variation is not insignificant, and we will examine this effect further
in a later article, but we choose to ignore it in the present analysis.
We therefore choose, for present purposes, to restrict our attention to
the profile of the rotation rate in the equatorial section of the Sun.

It is now necessary to introduce a PDF characterizing the
distribution of rotation rates over the entire equatorial section of
the convection zone (CZ). This may be formed from the PDF of equation
(3.1) as follows:
\begin{equation}
P(\nu|CZ) = \frac{1}{r_u - r_l} \int\limits_{r_l}^{r_u} {\rm d}r
            P(\nu|r,0) .
\end{equation}
We may verify that
\begin{equation}
\int\limits_{\nu_a}^{\nu_b} {\rm d}\nu P(\nu|CZ) = 1 .
\end{equation}

We denote by $r_l$ and $r_u$ the lower and upper limits of the
convection zone. The former is set by the upper limit of the
tachocline, so $r_l = 0.74$, and the latter by the photosphere, so $r_u
= 1$. Over this range the minimum sidereal frequency is $14.27 \,{\rm
y}^{-1}$, and the maximum is $14.83 \,{\rm y}^{-1}$, a range of almost
4\%. The resulting PDF is shown in Figure 3.

\begin{figure}[t]
\epsfig{file=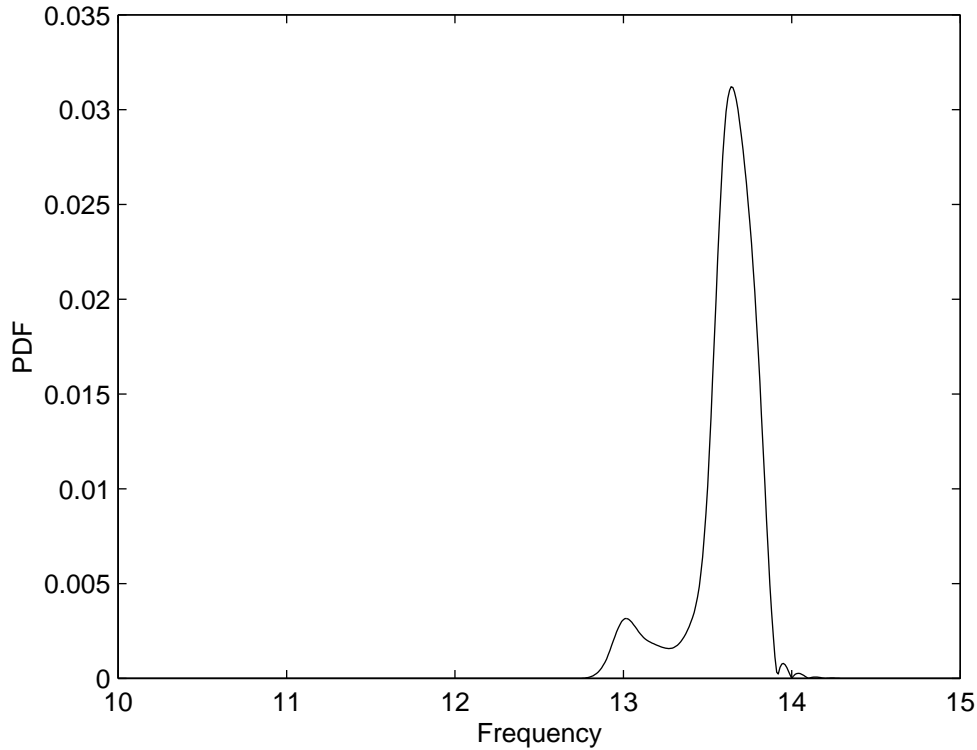}
\caption{This probability distribution function is representative of the
         range of (synodic) rotation frequencies in the convection zone.}
\end{figure}

We now form the ``integral resonance statistic'' 
\begin{equation}
\Gamma = \int\limits_{\nu_a}^{\nu_b} {\rm d}\nu S(\nu) P(\nu|CZ) ,
\end{equation}
where $S(\nu)$ is the power spectrum computed by the Lomb-Scargle
process in Section 2.  [This procedure is known in Bayesian parlance
(see, for instance, Bretthorst 1988) as an integration over a
``nuisance parameter.''] We denote by $\Gamma_d$ the actual value of
$\Gamma$ derived from the data, and we find that $\Gamma_d = 2.762$.
We now need to assess the significance of this value.

We adopt the method introduced by Bahcall and Press (1991) in their
study of the apparent anticorrelation between Homestake measurements
and the sunspot number. The procedure is to ``shuffle'' the data many
times, reassigning the flux measurements among runs.  We have carried
out 10,000 shuffles of the data and the results are shown in Figure 4.
We find that, of the 10,000 random simulations, only 16 have values
$\Gamma > \Gamma_d$.  From this analysis, we infer that there is a
probability of less than 0.2\% of obtaining the actual value of
$\Gamma$ by chance.  We conclude that the modulation of the neutrino
flux, as measured by the GALLEX-GNO experiment, in the frequency band
corresponding to the solar convection zone, is significant at the 0.2\%
level.

\begin{figure}[t]
\epsfig{file=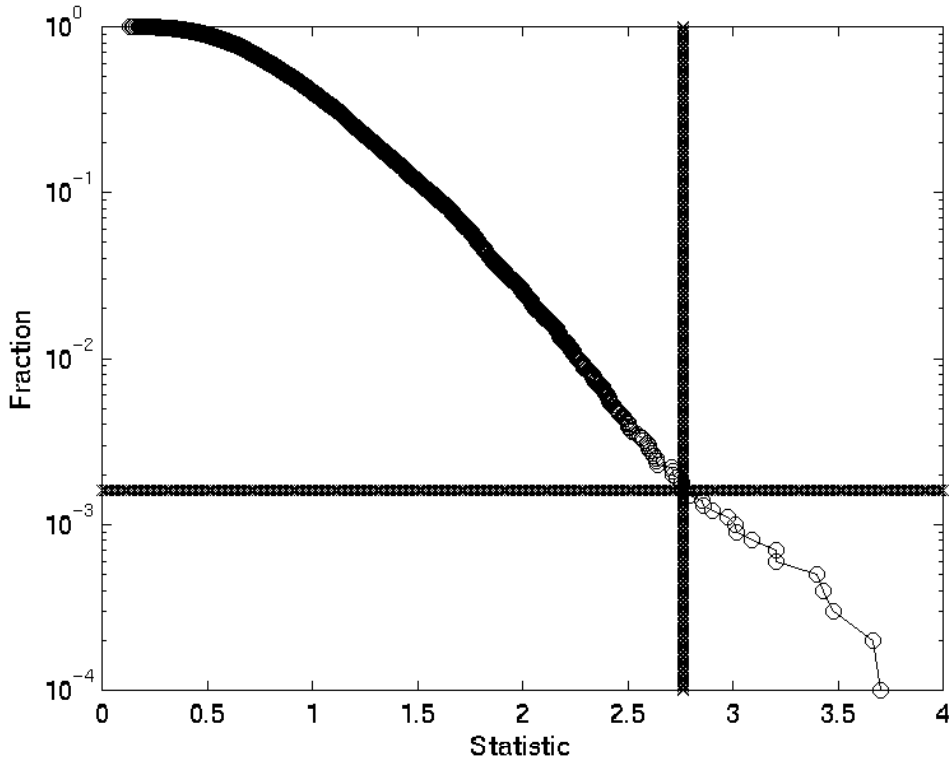}
\caption{For analysis of the convection zone, the ordinate denotes the 
         fraction of 10,000 simulations that have values of the statistic 
         $\Gamma$, defined by equation (4.3), larger than the value given 
         by the abscissa. The vertical line denotes the actual value of 
         the statistic $\Gamma_d$, derived from the data. Less than 0.2\% 
         of the simulations have $\Gamma > \Gamma_d$.}
\end{figure}

\section{Test of Possible Radiative-Zone Modulation}

We have repeated the calculations made in Section 4, applying them to
the radiative zone (RZ). We use equation (4.1) to calculate
$P(\nu|RZ)$, except that we now adopt $r_l = 0.30$ and $r_u = 0.66$.
The resulting PDF is shown in Figure~5.

\begin{figure}[t]
\epsfig{file=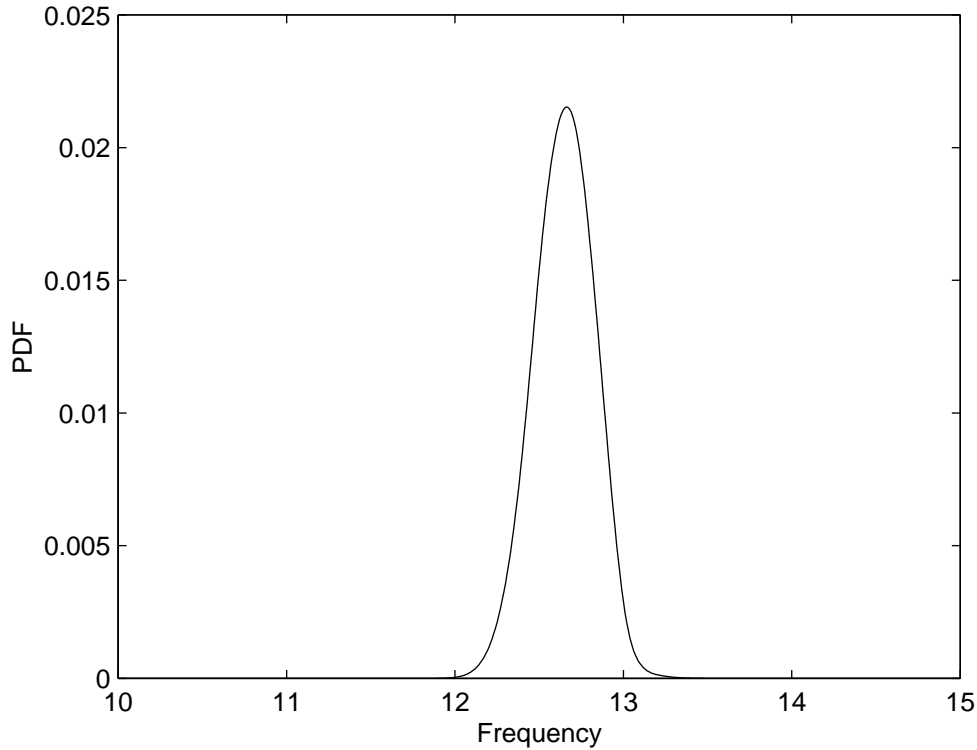}
\caption{This probability distribution function is representative of the 
         range of (synodic) rotation frequencies in the radiative zone.}
\end{figure}

When we calculate the resonance statistic $\Gamma$, using equation
(4.3), we obtain the value $\Gamma_d = 1.140$. We have carried out
1,000 shuffles of the data, and the results are shown in Figure 6.  We
see that, of the 1,000 random simulations, almost 300 have values
$\Gamma > \Gamma_d$.  From this analysis, we infer that there is no
significant resonance between oscillations of the solar neutrino flux
and rotation of the radiative zone.

\begin{figure}[t]
\epsfig{file=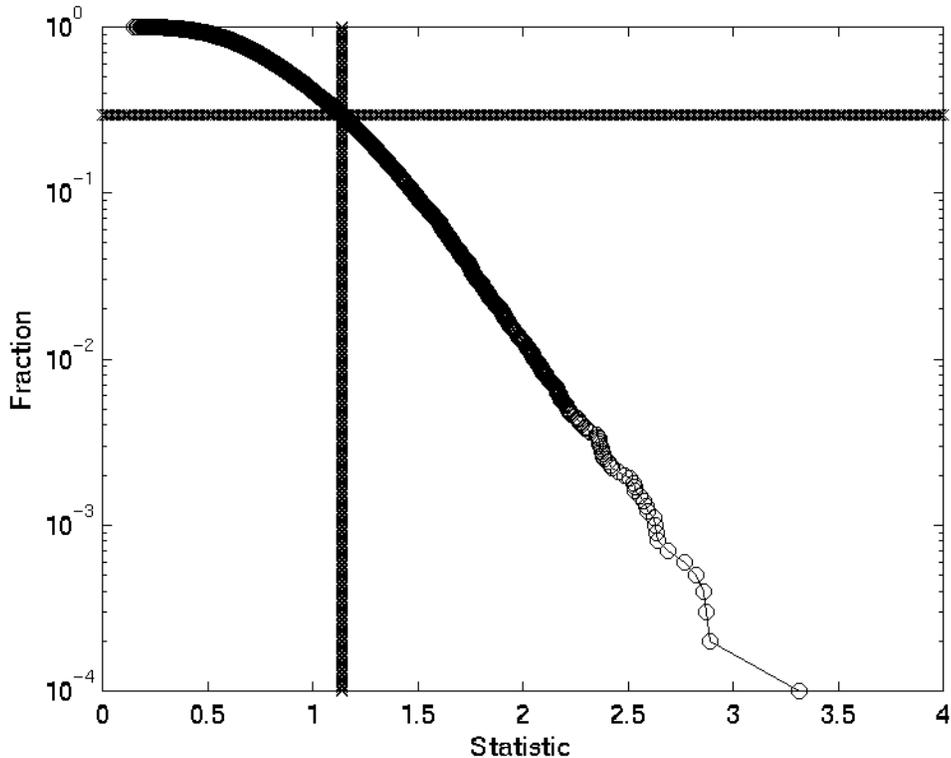}
\caption{For analysis of the radiative zone, the ordinate denotes the 
         fraction of 1,000 simulations that have values of the statistic 
         $\Gamma$, defined by equation (4.3), larger than the value given 
         by the abscissa. The vertical line denotes the actual value of 
         the statistic $\Gamma_d$, derived from the data. By contrast with 
         the situation shown in Figure 4, almost 30\% of the simulations 
         have $\Gamma > \Gamma_d$.}
\end{figure}

\section{Discussion}

We shall continue to examine the evidence for rotational or other
modulation of the solar neutrino flux using independent tests and, as
far as this is possible, independent data sets. We will certainly
repeat the resonance-statistic analysis for the Homestake (Davis \& Cox
1991; Lande et al.\ 1992; Cleveland et al.\ 1995, 1998) and SAGE
(Abdurashitov et al.\ 1999) data, since these data sets are publicly
available. We will also carry out a similar analysis of the SNO
(Sudbury Neutrino Observatory) data when in due course those data
become available, and of the Kamiokande and/or Super-Kamiokande data in
the happy event that those data are one day released to the public.
Since the GONG (Global Oscillations Network, National Solar
Observatory) consortium has recently carried out a new analysis of
solar internal rotation, we plan to repeat the previous calculation
using the results of that analysis.

As we mentioned in the introduction, spectrum analysis of radiochemical
data requires great care. We believe that the Lomb-Scargle analysis as
used in this article is trustworthy, but this is not to say that other
procedures may not yield superior information.  To this end, we plan to
develop a modification of the Lomb-Scargle procedure that takes account
of the data acquisition process, including the exponential decay of
capture products, inherent in radiochemical experiments. This will be
similar to, but probably not identical to, the maximum-likelihood
procedure that we used to analyze the Homestake data (Sturrock et
al.\ 1997).

In view of the results of this article, it will be interesting to
review the claimed correlations between neutrino measurements and
familiar solar indices such as sunspot number, surface magnetic field
strength, etc., which we discussed in Section 1. If the rotational
modulation of the neutrino flux is real, as it now appears to be, this
could provide an explanation of these correlations, since almost every
solar index also displays such a modulation. If Homestake data were the
basis of an earlier study , we will repeat the analysis using
GALLEX-GNO data. If it is possible to use an alternative and
independent source of data for a solar index, we shall do so. If a new
study shows evidence of a significant correlation between the solar
neutrino flux and another solar index, we will test to see if that
correlation is due partly or completely to the fact that they are both
influenced by rotational modulation.

If we accept the evidence that modulation of the solar neutrino flux,
as measured by the GALLEX-GNO experiment, occurs in the low convection
zone, and if we assume that the modulation is due to a known mechanism
such as the VVO or RSFP process, we can begin to estimate the neutrino
parameters and magnetic field strength required to explain this result.
Once these parameters are estimated, we should be in a position to
predict whether or not similar modulation should be detectable by
high-energy experiments such as Kamiokande, Super-Kamiokande, and SNO.

\section{Acknowledgments}

This article is based on work
supported in part by NAS grants NSS 8-37334 and NAG5-4038, and NSF
grant ATM-9910215. It is a pleasure to acknowledge the interest of and
helpful suggestions from Evgeni Akhmedov, Blas Cabrera, Sasha
Kosovichev, John Leibacher, Joao Pulido, Jeff Scargle, Guenther
Walther, and Mike Wheatland.

\end{document}